\pdfoutput=1

\documentclass[a4paper,conference]{IEEEtran}
\pdfoutput=1

\ifCLASSINFOpdf
\usepackage[pdftex]{graphicx}
\else
\fi
\ifCLASSOPTIONcompsoc
  \usepackage[caption=false,font=normalsize,labelfont=sf,textfont=sf]{subfig}
\else
  \usepackage[caption=false,font=footnotesize]{subfig}
\fi
\usepackage{fixltx2e}

\usepackage{stfloats}
\usepackage{amsfonts}
\usepackage{multirow}
\usepackage{multicol}
\usepackage[lined,boxed]{algorithm2e}

\newcommand*\rot{\rotatebox{90}}

\title{A statistically constrained internal method for single image super-resolution}

\author{\IEEEauthorblockN{Pierrick Chatillon}
\IEEEauthorblockA{pierrick.chatillon@onera.fr\\
DOTA, ONERA, Université Paris Saclay\\
F-91123 Palaiseau-France,\\
LTCI, Télécom Paris, IP Paris}
\and
\IEEEauthorblockN{Yann Gousseau}
\IEEEauthorblockA{yann.gousseau@telecom-paris.fr\\
LTCI, Télécom Paris, IP Paris \\
19 place Marguerite Perey\\ 91120 Palaiseau, France}
\and
\IEEEauthorblockN{Sidonie Lefebvre}
\IEEEauthorblockA{sidonie.lefebvre@onera.fr\\
DOTA, ONERA, Université Paris Saclay\\
F-91123 Palaiseau-France
}}

\begin{document}
\maketitle

\begin{abstract}
Deep learning based methods for single-image super-resolution (SR)
have drawn a lot of attention lately. In particular, various papers
have shown that the learning stage can be
performed on a single image, resulting in the so-called internal
approaches. The SinGAN method is one of these contributions, where the
distribution of image patches is learnt on the image at hand and
propagated at finer scales. Now, there are  situations where some
statistical a priori can be assumed for the final image. In particular, many natural phenomena yield images having power law Fourier spectrum, such as clouds and other texture images.  
In this work, we show how such a priori information can be integrated into an internal
super-resolution approach, by constraining the learned up-sampling
procedure of SinGAN. We consider various types of constraints, related
to the Fourier power spectrum, the color histograms and the
consistency of the upsampling scheme. We demonstrate on various
experiments that these constraints are indeed satisfied, but also that
some perceptual quality measures can be improved by the proposed
approach.

\end{abstract}

\graphicspath{ {./images/} }
\maketitle
\section{Introduction}

Single image super-resolution has recently benefited from tremendous progresses thanks to the use of convolutional neural networks, 
from the early SRCNN architecture \cite{dong2015image}, through many methods improving the PSNR of SR, including EDSR \cite{EDSR}, RDN \cite{RDN} or the recent state-of-the-art work SwinIR~\cite{liang2021swinir}. Such methods usually rely on an extensive 
training on generic image databases and are sometimes referred to as {\it external methods}. More recently, it has been shown that the mere structure of convolutional neural networks provides a good enough 
prior for several  image restoration tasks \cite{ulyanov2018deep}. Further, and taking advantage of the learning power of Generative Adversarial Networks (GAN), 
methods have been proposed to train a super-resolution architecture on a single image, most notably through the SinGAN approach~\cite{SinGAN}. Such methods are often coined as {\it internal methods}, and can be seen as an extension to neural networks of the general principles underlying patch-based approaches exploiting the statistical property of patches from the input image, such as \cite{singleImageSR} or \cite{internal_stats_single_image}. 
While being slightly less efficient than external methods, internal methods have the advantage of not necessitating an auxiliary image database that should have the same statistical property
as the images to be restored. 

In this work, we take interest in the application of internal super-resolution methods to a specific type of images, namely "colored noise" images that is, images for which the power spectrum decays as a power function $1/f^\beta$. A common such example is pink noise, also known as $1/f$ noise. While such a  decay is a general property of natural images~\cite{field1987relations},
we here consider images for which this behavior is particularly accurate and isotropic. Colored noise processes are particularly 
abundant in nature~\cite{keshner19821}, and can be encountered, as far as dimension two is concerned, in images of fracture patterns~\cite{hirata1987fractal}, glacier surface~\cite{arnold2003self} and of clouds, which are our main application cases. Such processes have also been used as a simple yet very efficient source of textures in computer graphic applications~\cite{perlin1985image}.

\begin{figure}[t]
    \centering
    \includegraphics[width=.99\linewidth]{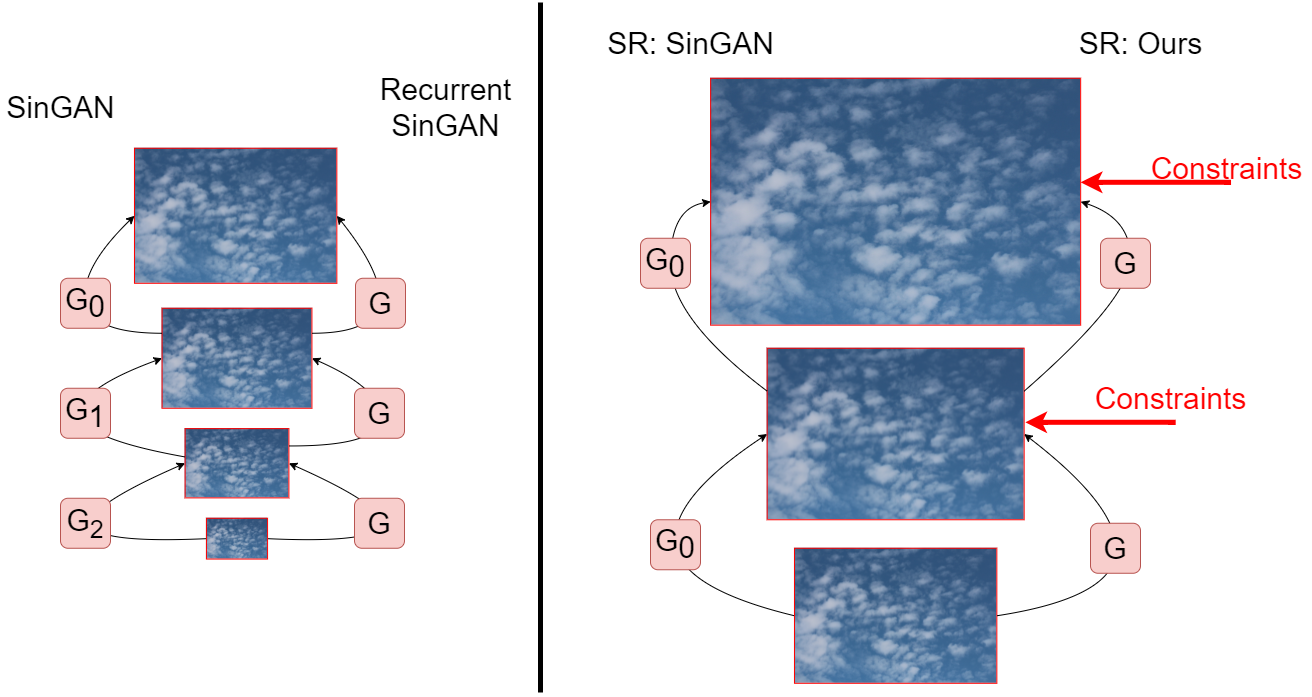}
    \caption{\centering Overview of the training (left) and SR procedure (right) for SinGAN and recurrent SinGAN \cite{RSinGAN}. This illustrates how we stabilize SinGAN's SR method by adding statistical constraints.} 
    \label{schema}
\end{figure}

Generating large, realistic and high resolution  cloud fields is important either in the context of evaluating the performance of optics sensors or to benchmark radiative transfer models~\cite{venema}. Turbulent flows simulations, such as the large-eddy ones, are expensive and stochastic models, based on fractal or Fourier framework, have been proposed in order to simulate more quickly realistic cloud fields. Examples include the bounded cascade model~\cite{cahalan94}, the iterative amplitude adapted Fourier transform algorithm~\cite{venema}, and the tree-driven mass accumulation process (tdMAP) model~\cite{benassi}. They exploit the scale invariant properties observed in real clouds: the power spectra of the logarithm of their optical properties is characterized by a spectral slope of around $\frac{5}{3}$ ~\cite{cahalan89}~\cite{benassi}~\cite{szczap} from small scale (a few meters) to the outer scale (about one-hundred kilometers), where the spectrum becomes flat due to decorrelation. The goal of the present work is to develop an SR method in the spirit of these generative approaches, allowing the increase of resolution of colored noise images while preserving physically pertinent statistical property such as the power spectrum. To the best of our knowledge, such a model has not been proposed in the literature, except for ~\cite{chainais_soleil}, where a multi-fractal model is leveraged for the super-resolution of images from the quiet sun. 


We also argue that imposing statistical constraint to an SR method is a way to overcome the limitation of PSNR training/evaluation. As pointed out in \cite{tradeoff}, PSNR alone is not enough, and two kind of metrics actually compete against each other: distortion metrics (e.g. PSNR) and perception metrics (e.g. perceptual quality \cite{perceptual_loss}, NIQE \cite{niqe}).
Optimizing for PSNR often leads to smooth images, when a perceptually good image needs some sharpness. Following this principle, many methods try to produce realistic SR outputs, using the GAN framework (introduced in \cite{vanilla_GAN}): SRGAN \cite{SRGAN}, ESRGAN \cite{ESRGAN} and EnhanceNet\cite{EnhanceNet} combine the GAN training with distortion losses, perceptual quality losses and even texture aware losses \cite{gatys2015texture}.\\ In the experimental part of this paper, Section \ref{sec:expe}, we will show that the chosen constraints indeed improves such perceptual (NIQE) and texture metrics. 

\section{Method}
\subsection{Statistical constraints}
In this section, we describe the image statistics we impose to the SR framework.  Some of these constraints are related to the specific SR task, while others are specific to colored noise images and related to desired physical property of the result.

Throughout the article, we denote as $u$ the low resolution image (LR) and $w$ the result of super-resolution by a factor $f>1$. The notation $\xi$ is for the frequency variable in the Fourier domain.
\subsubsection{Spectral constraints}
As explained in the introduction, we are interested in images whose spectrum decreases as a power function. Moreover, it is known that quasi-stationary signals such as cloud images yield uncorrelated discrete Fourier transform coefficients \cite{rayleigh}, and as a consequence, that the modulus follows a Rayleigh distribution. Therefore, the spectrum of the generated image is constrained so that : 
\begin{itemize}
    \item  Past a radius $r_0$ in the Fourier domain, the mean modulus of $\hat{w}$ taken on a circle $\mathcal{C}_{r}$ of radius r decreases like $\mathbb{E}_{\mathcal{C}_{r_0}}(\|\hat{w}\|) (\frac{r}{r_0})^{-p}$.\\
    \item Fourier coefficients with frequencies of equal modulus are Rayleigh-distributed.
\end{itemize}

To achieve both these objectives, we take the ordered moduli of all coefficients lying on a circle of radius $r$, then modify them so that they correspond to ordered samples from a Rayleigh distributed variable with the desired mean $\frac{\mathbb{E}_{\mathcal{C}_{r_0}}}{r^p}$.\newline
We write $\xi^r_i$ for the $|\mathcal{C}_{r}|$ frequencies lying on $\mathcal{C}_r$ ordered by magnitude of $\|\hat{w}_\xi\|$, and  $Q^{\mathcal{R}}_{\beta}$ for the quantile function of the Rayleigh distribution of parameter $\beta$: $$Q^{\mathcal{R}}_{\beta}(u)= \beta \sqrt{-2 \log(1-u)}$$
Finally, we set $\beta_r$ such that the modulus has the desired mean over $\mathcal{C}_r$: $\beta_r = \sqrt{\frac{2}{\pi}} \frac{\mathbb{E}_{\mathcal{C}_{r_0}}(\|\hat{w}\|)}{\mathbb{E}_{\mathcal{C}_{r}}(\|\hat{w}\|)} (\frac{r}{r_0})^{-p} $ . Eventually, the spectrum is imposed using the  projection $proj_{spectrum}$, defined as:
\begin{equation}
\left\{
    \begin{array}{cccl}
        proj_{spectrum}(\hat{w}) &=& \hat{w_\xi} &\mbox{ if } r \leq r_0 \\
        proj_{spectrum}(\hat{w})_{\xi^r_i} &=& \frac{\hat{w}_{\xi_i^r}}{\|\hat{w}_{\xi_i^r}\|} Q^{\mathcal{R}}_{\beta_r}(\frac{i}{|\mathcal{C}_{r}|}) &\mbox{ if } r > r_0
    \end{array}
\right.
\label{proj_slope_rayleigh}
\end{equation}

\begin{figure*}[t]
    \centering
    \subfloat[Ground truth]{\includegraphics[width=.33\linewidth]{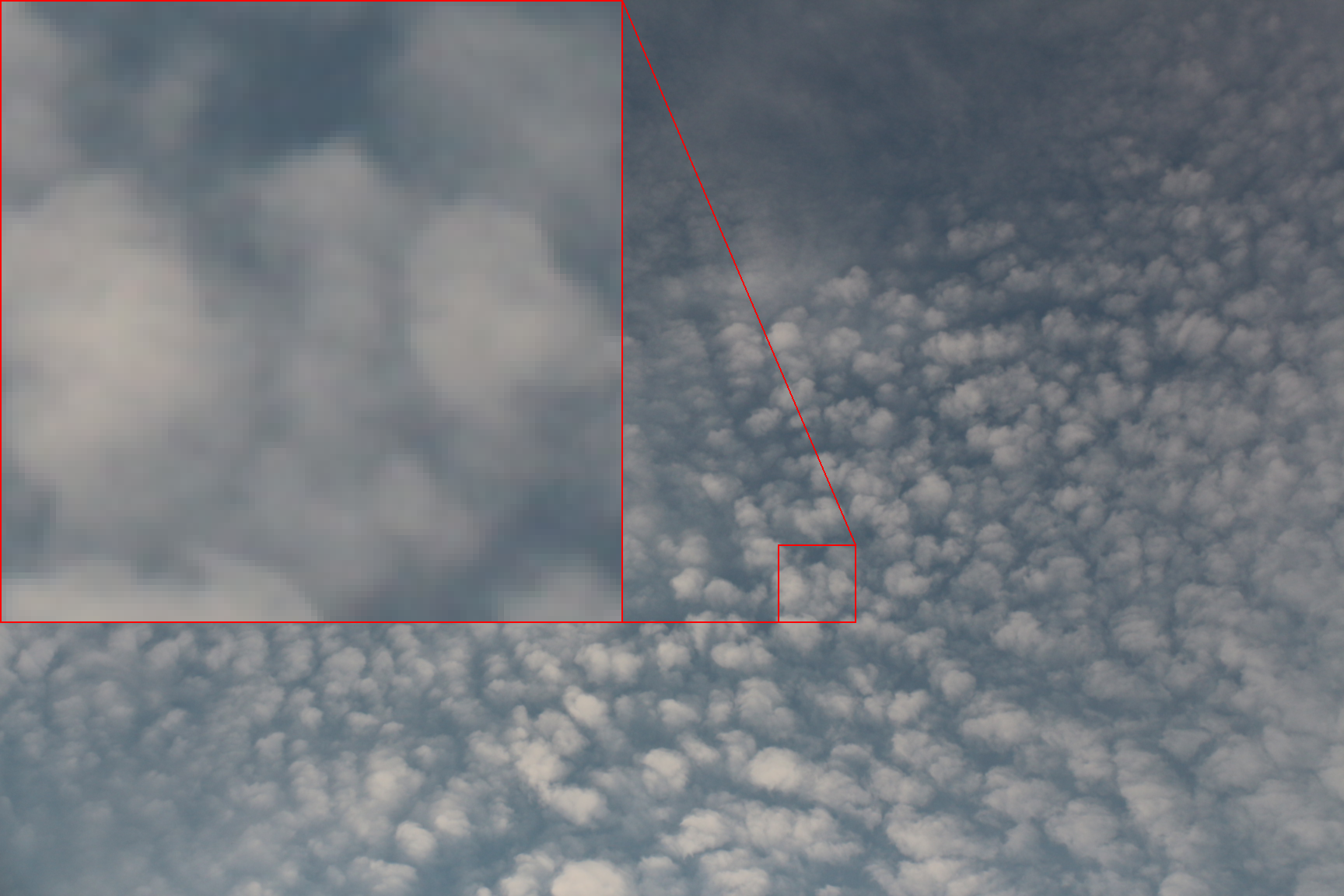}}
    \subfloat[Bilinear interpolation]{\includegraphics[width=.33\linewidth]{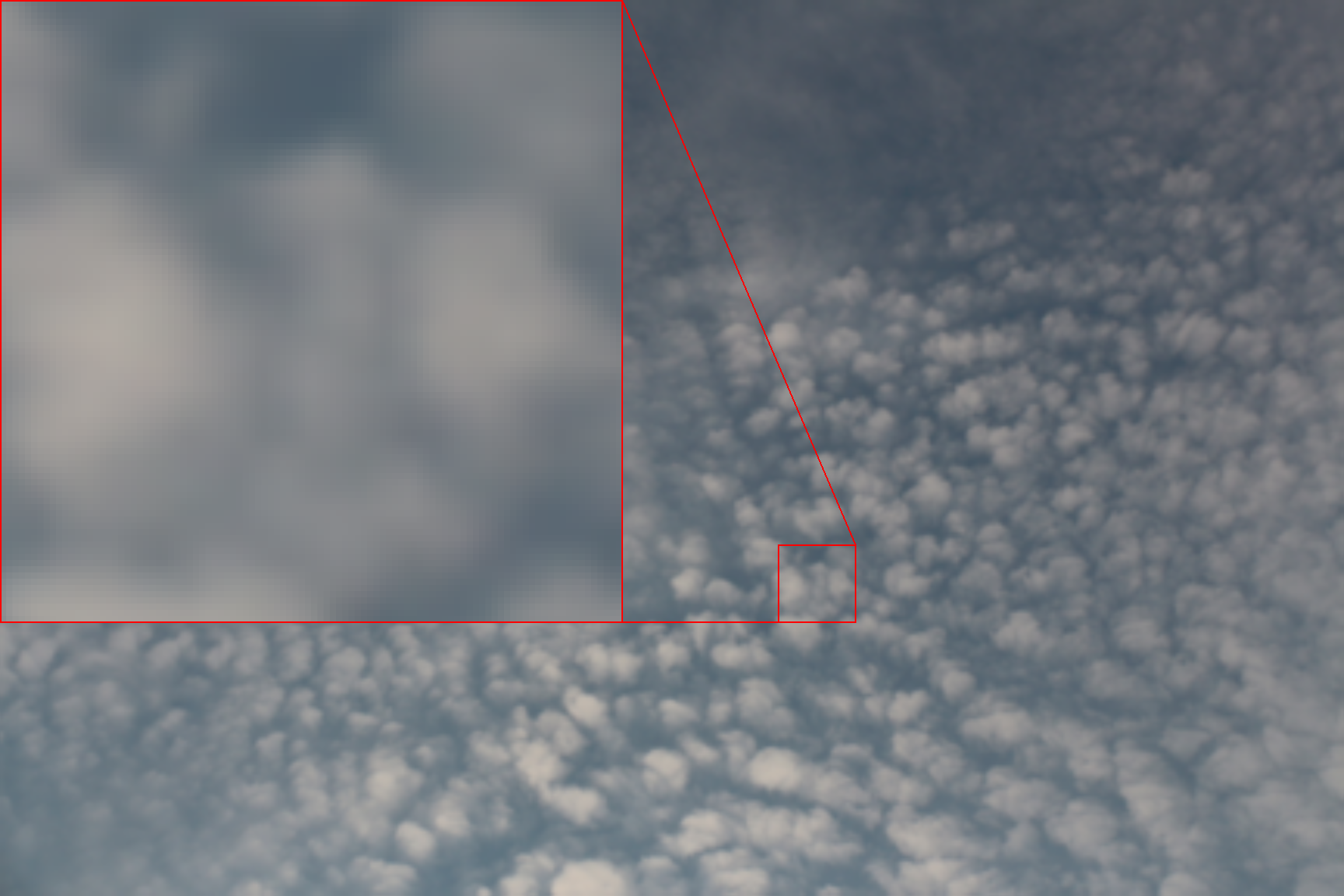}}
    \subfloat[RDN]{\includegraphics[width=.33\linewidth]{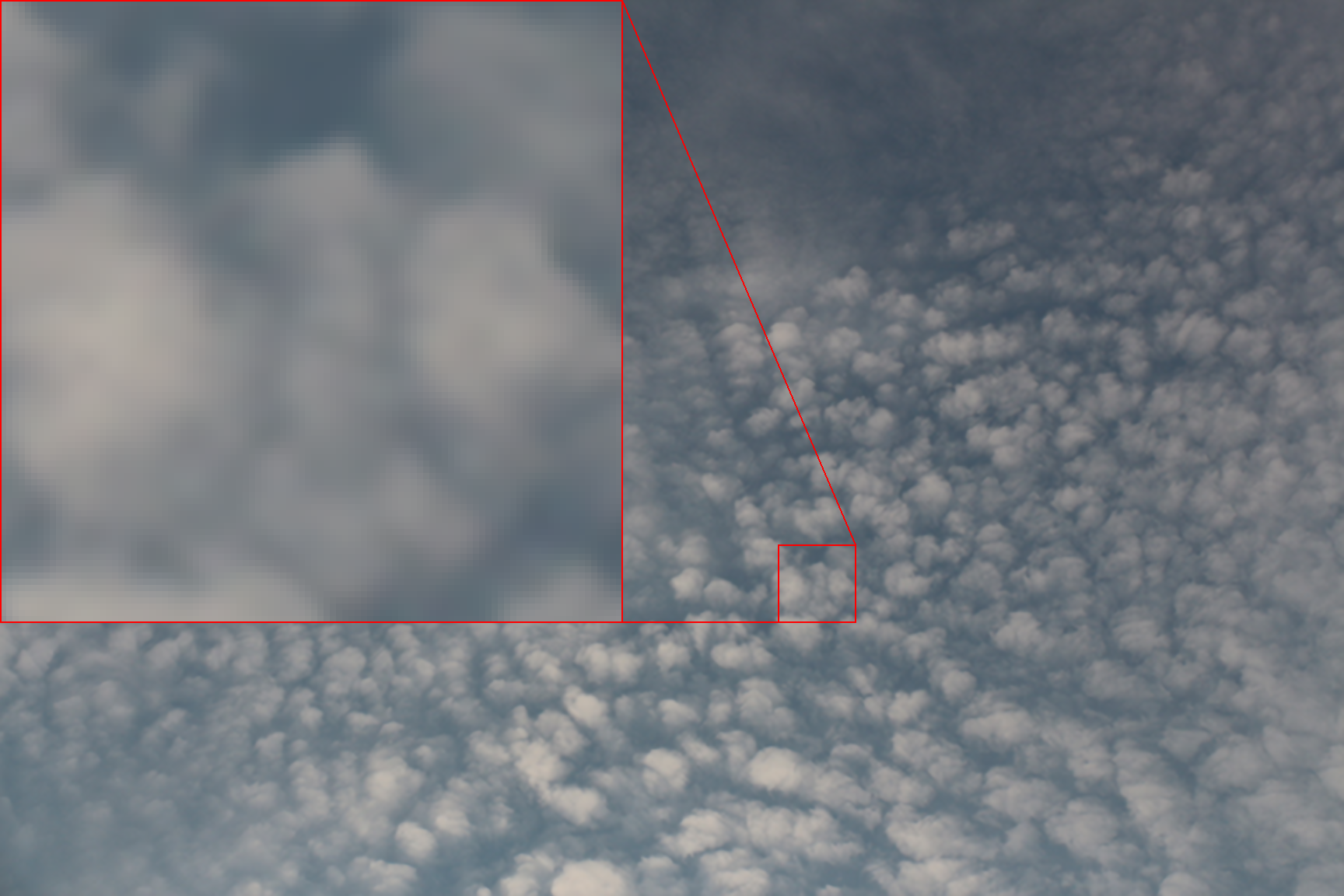}}\\
    \subfloat[SinGAN]{\includegraphics[width=.33\linewidth]{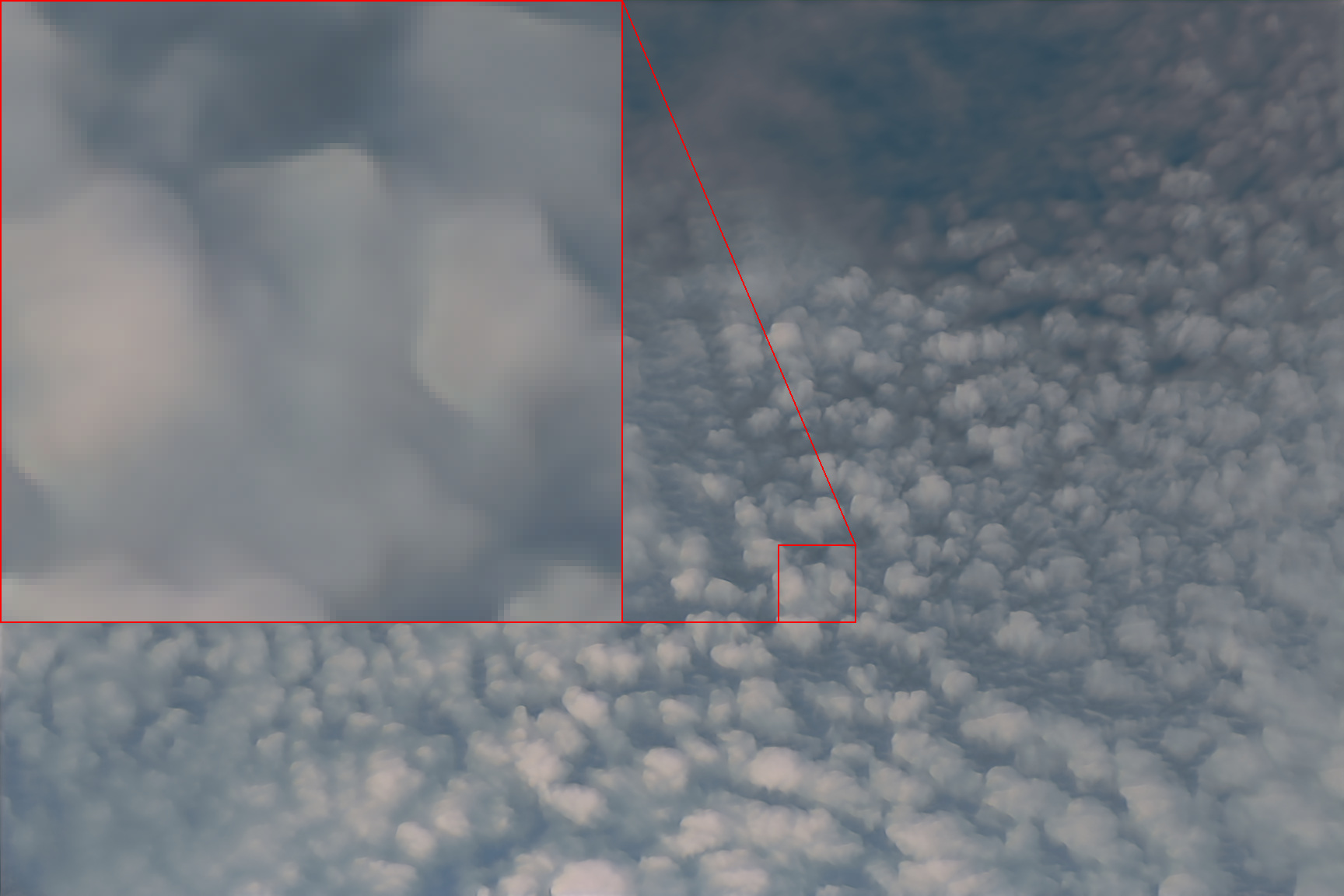}}
    \subfloat[Ours, with constraints]{\includegraphics[width=.33\linewidth]{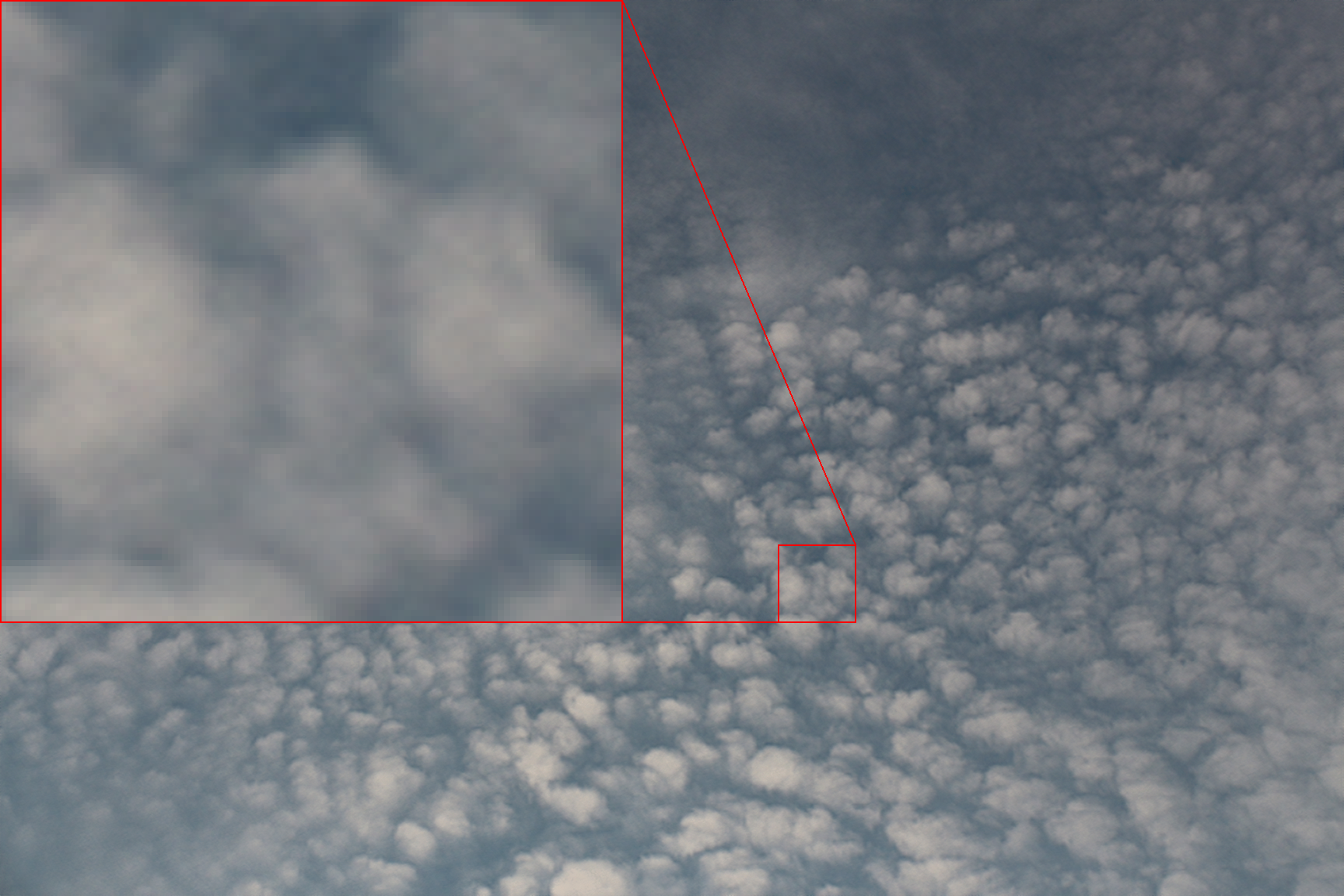}}
    \subfloat[Ours, without constraints]{\includegraphics[width=.33\linewidth]{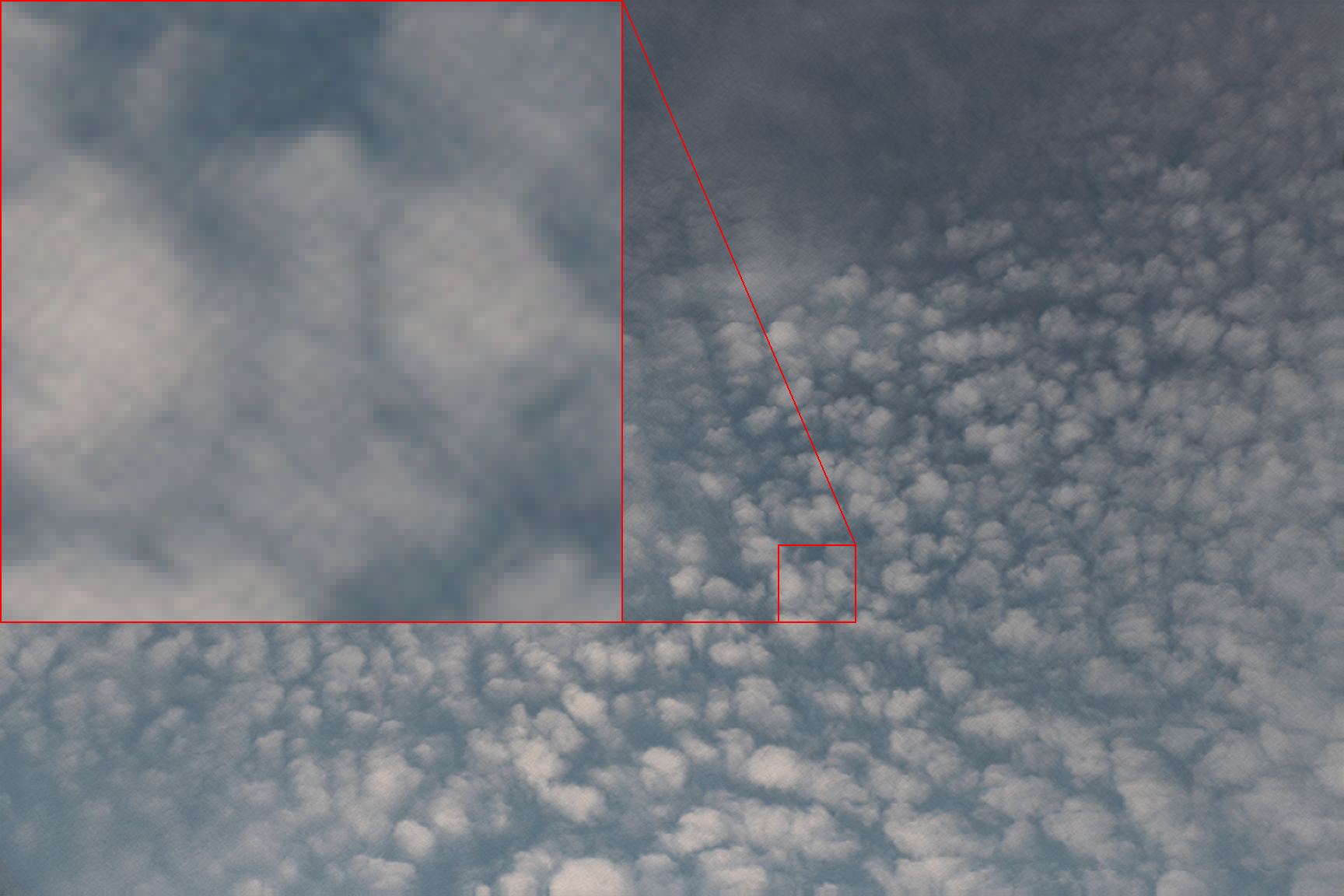}}

    \caption{\centering Visual comparison of all studied methods for an image taken from our Cloud pictures dataset. The highlighted zone is 100*100 pixels.} 
    \label{cropcomp}
\end{figure*}
Additionally, to avoid artifacts and control only the relevant part of the spectrum, we rely on the 'periodic plus smooth decomposition' of an image \cite{Moisan2011}. This article proposes to get rid of spurious horizontal and vertical structures in the Fourier domain, induced by the implicit periodization of the image during the Fourier transform. The approach consists in decomposing the image into the sum of a smooth non-periodic image, and a periodized version of the image that contains most of the initial information.\\
Following the author's guideline, we perform our spectrum operations on the periodic image before adding back the untouched smooth image.

\subsubsection{Histogram constraint}
In order to counteract some possible limitations of the generator, we want the color histogram of $w$ to match $u$'s.
We adopt a sliced optimal transport approach \cite{rabin2011wasserstein}.\\
This approach consists in iterating projection steps:
for each step, we sample a random unit vector $\theta$ of the color space. We then modify the color histogram of $w$ along the direction $\theta$ such that the histograms projected onto $\theta$ of $u$ and $w$ are closer.\newline
Each step reads: $w_{\sigma^{w}_k} = w_{\sigma^{w}_k} + \epsilon \theta\cdot\langle u'_{\sigma^{u'}_k}-w_{\sigma^{w}_k}, \theta \rangle$ ,\newline
where $u'$ is a version of $u$ that has been stretched to match the size of $w$, and $\sigma^{image}$ is the permutation that rearranges $\langle image,\theta \rangle$ in increasing order.\newline
We call $proj_{hist}(w)$ the result of this process.

\subsubsection{Reversibility constraint}
The reversibility condition is a consistency test between the input low resolution image $u$ and the super-resolution result $w$:\\
we want $u$ to be the result of blurring then subsampling $w$.\\
This condition can be seen as a fidelity term, making the upsampled result closer in the $L_{2}$ sense to the ground truth. In other words, we want our method to be balanced in the perception-distortion tradeoff (\cite{tradeoff}). This condition can be computed as the following loss:
\begin{equation}
\mathcal{L}_{rev}= \frac{1}{2}\| D_f(w\ast g) - u \|^2
\label{loss_rev}
\end{equation}
where $D_f$ is the decimation operator (from the dimension of $w$ to the dimension of $u$), and the smoothing filter $g$ is chosen as a gaussian filter of standard deviation $\sigma=.7*f$, which is a reasonable filter size regarding Shannon's theorem.
We call the resulting image $proj_{rev}(w)$.

\subsection{SinGAN}

In order to be able to adapt to the statistical specificity of a single image, we chose as a backbone the classical SinGAN architecture. We first quickly recall the architecture of this network and the way it can be applied to SR.  

\subsubsection{Architecture}
First, a single training image is turned into a multi-resolution pyramid $\{x_N, ..., x_0\}$ where $x_0$ is the original image, and $x_n$ is $x_0$ downsampled $n$ times by a factor $r>1$.\\
SinGAN is based on a pyramidal architecture of generators $\{G_N, ..., G_0\}$ and of discriminators $\{D_N,...,D_0\}$, each generator $G_i$ being trained with the help of a patch-discriminator \cite{pix2pix} $D_i$ in the WGAN-GP framework \cite{WGAN-GP}.\\
The coarsest scale generator $G_N$ is unconditional and maps a Gaussian noise $z_N$ onto a plausible low-resolution image $\tilde{x_N}$.\\
Then, at each step $n$ of the pyramid, the same operation is performed by the residual generator $G_n$: $\tilde{x}_{n+1}$ is upsampled by a factor r ($(\tilde{x}_{n+1})\uparrow ^r$) and some noise $z_n$ is added. The result is fed through the generator, which outputs details that are added to $(\tilde{x}_{n+1})\uparrow ^r$, to get $\tilde{x}_n$.\\
As the authors put it, if $\psi_n$ is the convolutional part of the generator $G_n$, then $G_n$ performs:
\begin{equation}
    \tilde{x}_{n}=\left(\tilde{x}_{n+1}\right) \uparrow^{r}+\psi_{n}\left(z_{n}+\left(\tilde{x}_{n+1}\right) \uparrow^{r}\right)
\end{equation}


The goal of the generators $G_n$ is to learn the patch distribution of $x_n$.
Moreover, a reconstruction loss is added, and weighted by a parameter $\alpha$.\\
The total loss reads: 
\begin{equation}
    \min _{G_n} \max _{D_{n}} \mathcal{L}_{\mathrm{adv}}\left(G_n, D_{n}\right)+\alpha \mathcal{L}_{\mathrm{rec}}\left(G_n\right)
\end{equation}
where $\mathcal{L}_{\mathrm{adv}}$ and $\mathcal{L}_{\mathrm{rec}}$ are described in SinGAN (\cite{SinGAN} Section 2.2).

\subsubsection{Super Resolution}
\begin{figure}[h!]
    \centering
    \subfloat[]{\includegraphics[height=.23\linewidth]{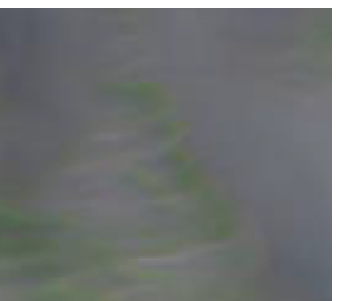}}
    \subfloat[]{\includegraphics[height=.23\linewidth]{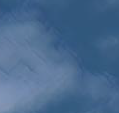}}
    \subfloat[]{\includegraphics[height=.23\linewidth]{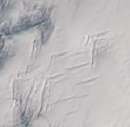}}
    \subfloat[]{\includegraphics[height=.23\linewidth]{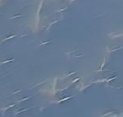}}
    \caption{Examples of SR artifacts using SinGAN on cloud images} 
    \label{artifacts}
\end{figure}

Once the finest resolution generator $G_0$ is trained (using a reconstruction weight $\alpha=100$ for SR), the authors propose to iterate this generator, yielding new image resolutions.\\ This step is arguably the weak point of SinGAN SR, and can yield strong color and directional patterns artifacts, as can be seen in the case of cloud images in Fig. \ref{artifacts}.\\

We argue that these artifacts come from the iteration of a network $G_0$ which has never been trained to receive its own output as input.
We tackle this issue in two ways:
\begin{itemize}
    \item We use a scale-invariant generator, as initially proposed in \cite{RSinGAN}
    \item We use the previously described constraints to the generation procedure, keeping some statistical properties while iteratively upsampling the image in a similar way to SinGAN's.
\end{itemize}

\subsection{Our approach}
 \subsubsection{Training procedure}
We train a unique residual generator G on the same pyramid as SinGAN, with 3 scales.
Since our aim is to upscale texture and not structure, we do not care for a deep multi-scale pyramid (which allows structural variability). Instead of generating a low resolution image with a GAN (as in SinGAN \cite{SinGAN} and Recurrent SinGAN \cite{RSinGAN}), we use a blurred and downsampled version of our training image as starting step. As a consequence, SinGAN uses as many generators as there are scales in the pyramid, Recurrent SinGAN uses 2, and we only use 1.\\
Next, following the principle of SinGAN, the image is iteratively passed through the generator, until original resolution is reached. In the original article SinGAN, the training takes place one generator at a time, from coarse resolution to high, because in order to train a given generator, we need to infer all the generators from coarser scales. This length of training is reduced by only training all stages of the pyramid at each training step. This is made possible by the fact that we use only one generator. As in SinGAN, we use a reconstruction loss, between the ground truth and the image obtained with our technique, using no noise.\\

\subsubsection{SR procedure}
In order to perform SinGAN-like SR under constraint, we alternate the application of our recurrent generator $G$ and 3 steps of alternate projection onto our 3 constraints as described in Algorithm \ref{altproj} and illustrated in Fig. \ref{schema}. This provides both a multi-scale upsampling in the same fashion as SinGAN, while respecting all constraints along the way.

\begin{algorithm}[h!]
\SetAlgoLined
\KwIn{$x_0$}
$w=x_0$\\
\While{desired zoom factor not reached}{
    $w=G(w)$\tcp{upscales the image, and adds detail}
    \For{step \textbf{in} range(3)}{
        $w=proj_{spectrum}(w)$\\
        $w=proj_{hist}(w)$\\
        $w=proj_{rev}(w)$
}}

\caption{\centering Alternate projection algorithm used to apply constraints to the image at each step of the iterative upsampling.}

\label{altproj}

\end{algorithm}

\subsubsection{Settings}
We choose the scaling factor used by \cite{SinGAN} for super resolution $r=0.793701$. We set the noise level $\sigma=0.01$ and $\alpha=100$. The Adam optimizer \cite{Adam} is used for the generator and the discriminators with learning rate $10^{-4}$, and parameters $(\beta_1,\beta_2)=(0.5,0.999)$. We train our network for 2000 iterations, with a learning rate decay of factor $0.1$ at the 1600$^{th}$ iteration (like SinGAN). 
At each GAN training step, the discriminator is trained for 3 iterations while the generator is only updated once, in the spirit of \cite{WGAN-GP}.

\begin{figure*}[t!]
\centering
\includegraphics[width=1.\linewidth]{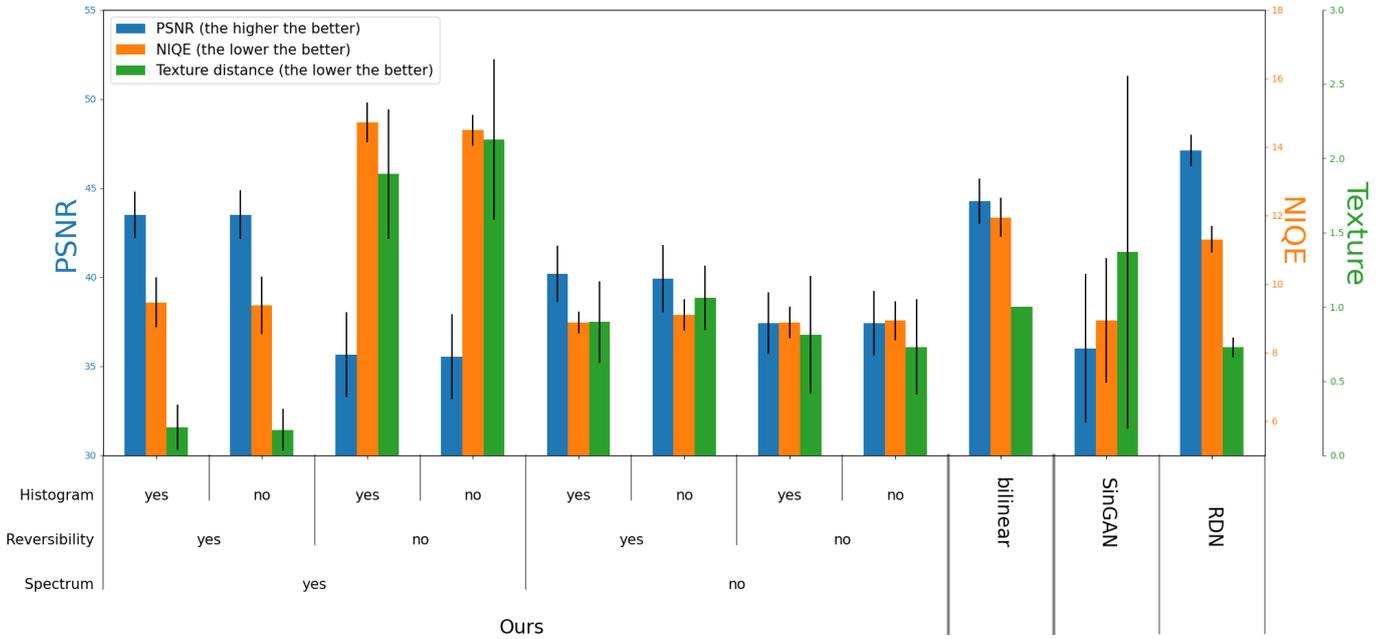}
\label{abla}
\caption{Ablation study realized on the whole Cloud dataset. Performances of bilinear interpolation, SinGAN and RDN are plotted for comparison purposes.}
\end{figure*}

\section{Experiments}
\label{sec:expe}
\subsection{Datasets}
\subsubsection{Synthetic dataset}
We create a synthetic dataset of 10 1200*1200 colored noise images. To forge them, we use the following method:
first, we sample a Gaussian noise image, then take its discrete Fourier Transform, and divide each coefficient by its modulus. This yields an array of unit complex numbers which we multiply by scalars having the desired power decay behavior.\\
We set the the exponent of this power decay to be $\beta=1.7$, approximating the theoretical exponent $\frac{5}{3}$ encountered in cloud images.

\subsubsection{Cloud dataset}
We create our own cloud pictures dataset from 12 1728*1152 photos of clouds taken from the ground. 
\subsubsection{BSD100 dataset}
The BSD100 dataset is a classical natural images dataset, used by the authors of SinGAN to evaluate their SR method. We'll also use it to compare our results with theirs.

\subsection{Evaluation metrics}
We compare SR results using different quality indices: PSNR, NIQE \cite{niqe}, and $D_{texture}$.\\
The NIQE index uses the distribution features extracted on patches to assess the quality of an image. Contrarily to the PSNR, it is a no-reference measure.\\
As done in \cite{houdard_wass}, and inspired by the article \cite{gatys2015texture}, we define a texture distance $D_{GATYS}$:\\
given an image I, we compute VGG features $F^{l}_{c,i,j}\in\mathcal{R}^{C \times N_l \times M_l}$ from the layer $l=pool4$ (the deepest representation of the 5 used layers). Then we compute the Gram matrix of these features: $G^{l}_{c,d}=\frac{1}{N_{l} M_{l}}\sum_{i,j} F_{c,i,j}^{l} F_{d,i,j}^{l}$. Finally, $D_{GATYS}$ is the $L^2$ distance between Gram matrices from both images.
\begin{equation}
    D_{GATYS}(I,\hat{I})=\frac{1}{4} \sum\left(G^{l}-\hat{G}^{l}\right)^{2}
\end{equation}
This distance varies a lot in magnitude upon taking different images. In order to compute meaningful statistics across multiple images, we normalize this distance for each image by the distance $d(I_{GT},I_{bilinear})$ between the ground truth image $I_{GT}$ and the 4x bilinear zoom of the downscaled image $I_{bilinear}$:
\begin{equation}
    D_{texture}(I_{GT},I)=\frac{D_{GATYS}(I_{GT},I)}{D_{GATYS}(I_{GT},I_{bilinear})} 
\end{equation}
As a consequence this distance is 1 when evaluating the bilinear interpolation against the ground truth. This distance can be seen as a ratio of performance when compared to the baseline bilinear interpolation.

All the experiments are run for a 4x zoom.
We compare our results against the bilinear interpolation, and RDN \cite{RDN}, an external SR method aiming at maximizing the PSNR.
Some results can be visually compared in Fig. \ref{cropcomp}, with a 100*100 section of the image magnified by a factor 8.

\subsection{Ablation study}
\begin{figure}[h]
    \centering
    \includegraphics[width=.95\linewidth]{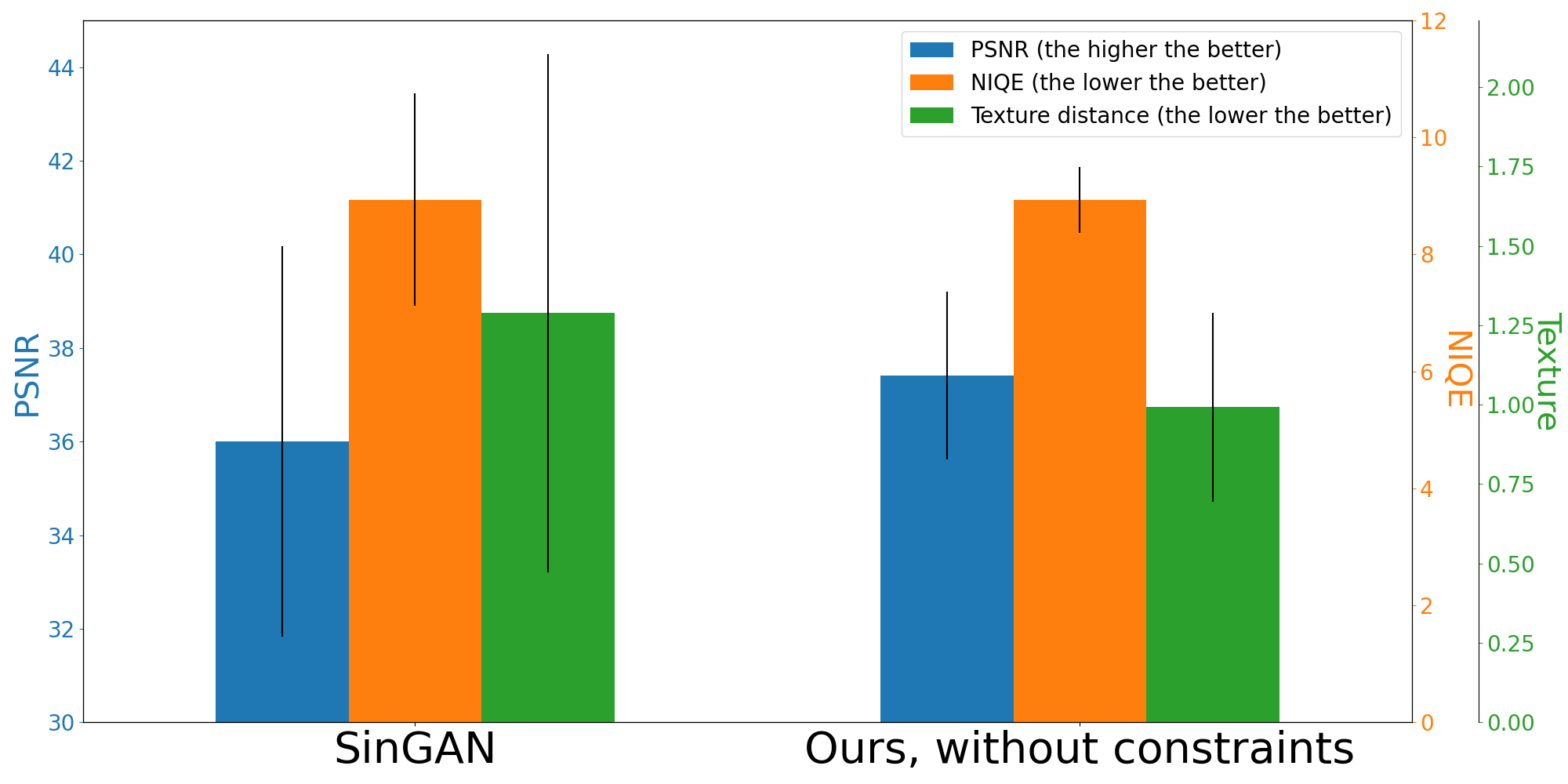}
   
    \caption{\centering Comparison between the original SinGAN and the use of a scale-invariant generator (without additional statistical constraints) on the Cloud dataset} 
    \label{no_constraints}
\end{figure}

First we verify that the choice of a recurrent generator helps to improve the results of SinGAN on auto-similar data, such as clouds. We notice indeed a slight improvement in PSNR and texture distance, when using our generator without any additional constraint (see Fig. \ref{no_constraints}). 

We then perform an ablation study on the various statistical constraints considered in this paper. Fig. 4 reports the performances of the method on the Cloud dataset when each constraint is used ('yes') or not ('no'). \\
We observe that, as expected, using the reversibility constraint improves the PSNR, and using both reversibility and spectrum yields the best texture score and PSNR of all choices.\\
Numerical results are reported in TABLE I.

\begin{table}
\centering
\begin{tabular}{|c|c|c|c|c|c|}

\cline{1-3}
\multirow{1}*{} \rot{Spectrum   }& \rot{Reversibility   } &  \rot{Histogram   }\\ 
\cline{4-6}
&&&PSNR & NIQE & $D_{texture}$ \\
\hline
\multirow{4}*{yes} & \multirow{2}*{yes} & yes & 43.5 $\pm$ 1.3& 9.5 $\pm$ 0.7 & 0.19 $\pm$ 0.15\\
\cline{3-6}
                                       & &  no & 43.5 $\pm$ 1.4 & 9.4 $\pm$ 0.8 & $\mathbf{0.17} \pm \mathbf{0.14}$\\
\cline{2-6}
                                       & \multirow{2}*{no} & yes & 35.6 $\pm$ 2.4& 14.7 $\pm$ 0.6& 1.89 $\pm$ 0.44  \\
\cline{3-6}
                                       &                    &  no & 35.5 $\pm$ 2.4 & 14.5 $\pm$ 0.4 & 2.13 $\pm$ 0.53 \\
\cline{1-6}
                    \multirow{4}*{no} & \multirow{2}*{yes} & yes &40.2 $\pm$ 1.6& $\mathbf{8.9}\pm  \mathbf{0.3}$& 0.90 $\pm$ 0.27 \\
\cline{3-6}
                                       &                    &  no & 39.9 $\pm$ 1.9& 9.1 $\pm$ 0.5 & 1.06 $\pm$ 0.22 \\
\cline{2-6}
                                       & \multirow{2}*{no} & yes & 37.4 $\pm$ 1.7 & $\mathbf{8.9}\pm  \mathbf{0.5}$ & 0.81 $\pm$ 0.40\\
\cline{3-6}
                                       &                    &  no &37.4 $\pm$ 1.8&$\mathbf{8.9}\pm  \mathbf{0.6}$ & 0.73 $\pm$ 0.32\\

\hline
 \multicolumn{3}{|c|}{Bilinear interpolation}&44.3 $\pm$ 1.3 & 11.9 $\pm$ 0.7 & 1 \\
\hline
 \multicolumn{3}{|c|}{SinGAN}&36.0 $\pm$ 4.1&$\mathbf{8.9}\pm  \mathbf{1.8}$ & 1.37 $\pm$ 1.18\\
 \hline
 \multicolumn{3}{|c|}{RDN}& $\mathbf{47.1} \pm \mathbf{0.9}$ & 11.3 $\pm$ 0.4 & 0.73 $\pm$ 0.07\\
\hline

\end{tabular}
\label{ablation_table}
\caption{Ablation results}
\end{table}

\subsection{Quantitative analysis}
We propose to compare bilinear interpolation, RDN, SinGAN, and our method using the metrics and datasets described above.

\begin{figure}[h!]
    \centering
    \subfloat[Comparison of different methods on our synthetic dataset]{\label{sanity_check}\includegraphics[width=.9\linewidth]{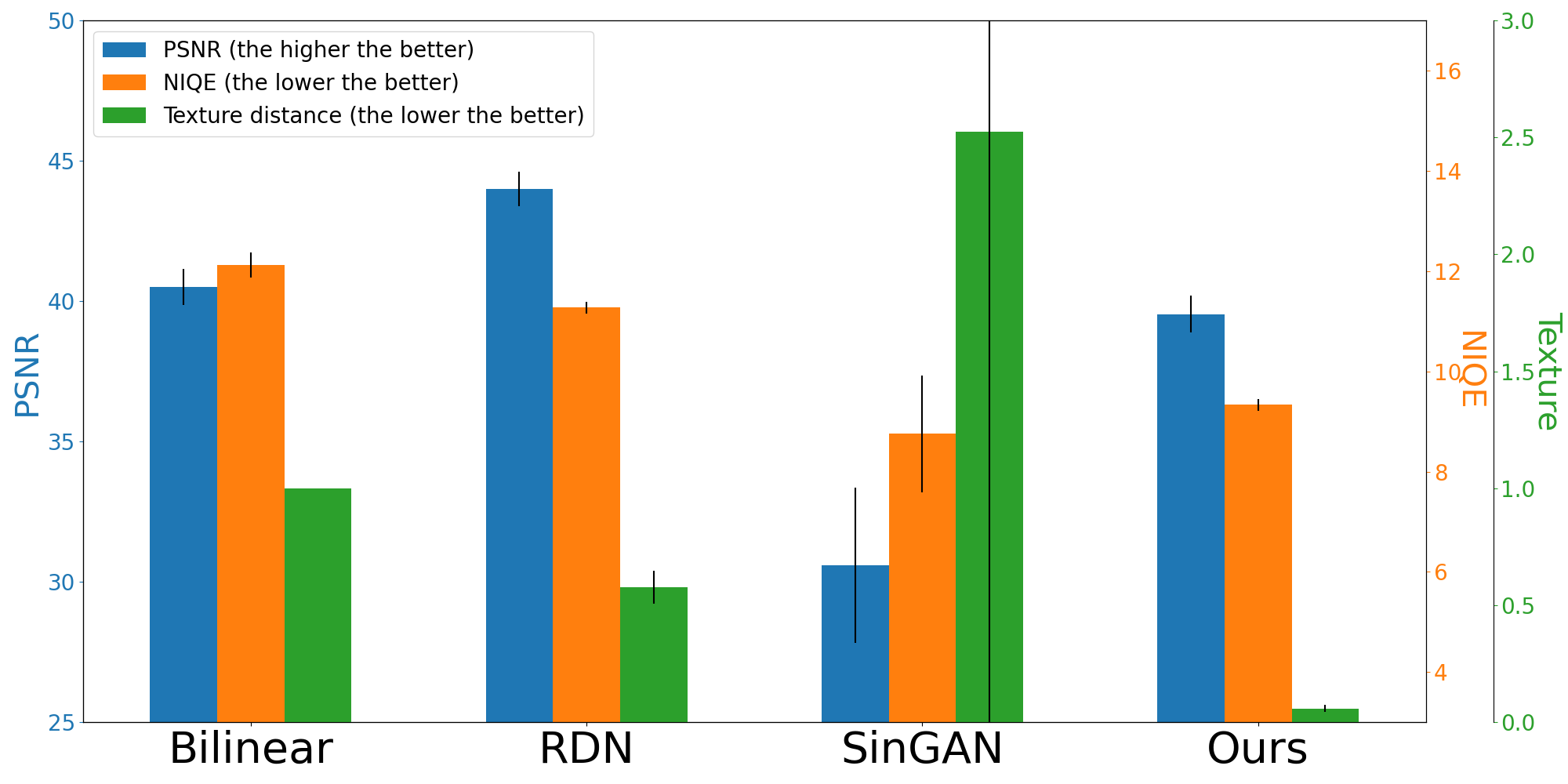}}

    \subfloat[Comparison of different methods on our Cloud dataset]{\label{eval_cloud}\includegraphics[width=.9\linewidth]{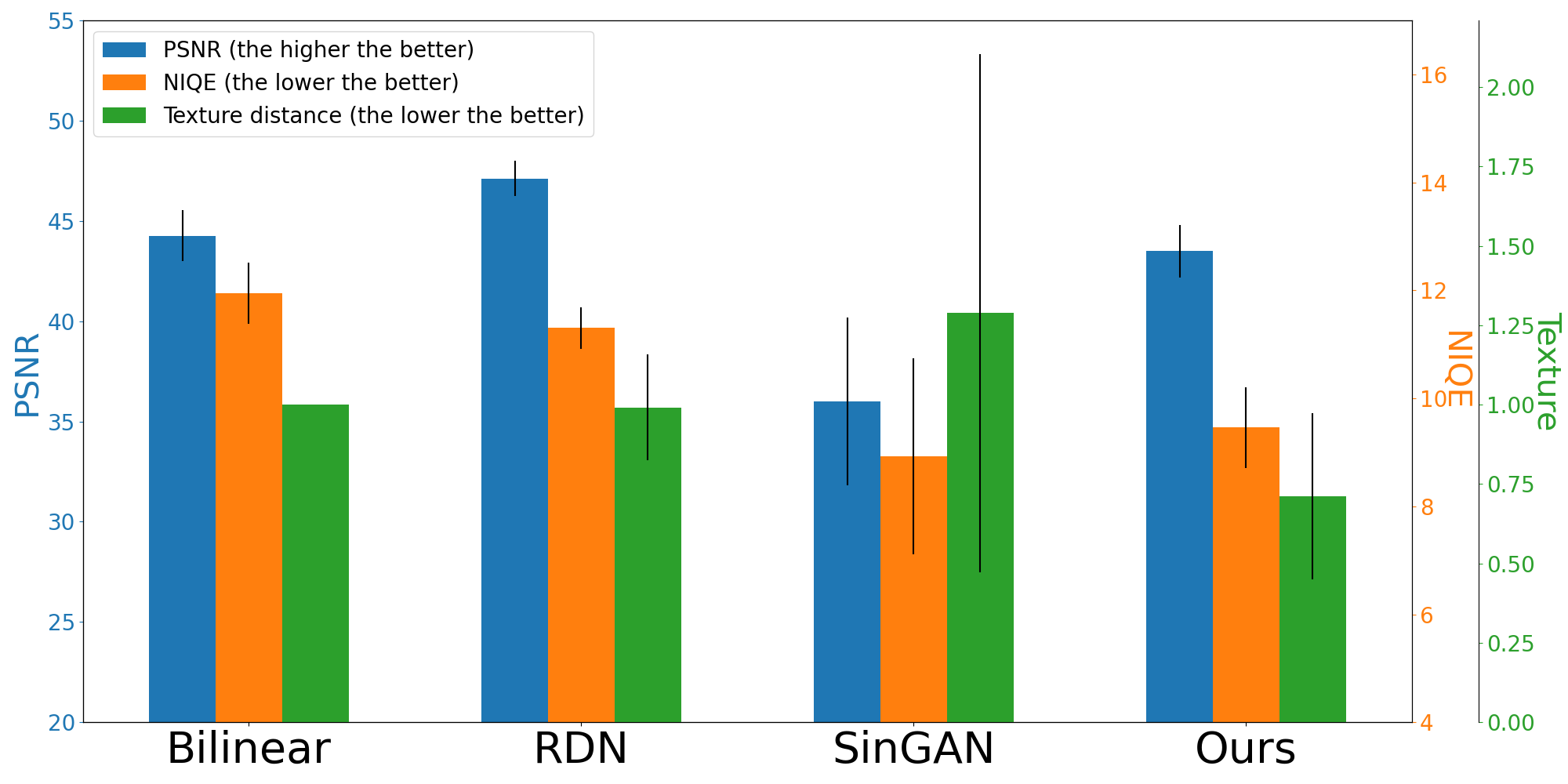}}
    
    \subfloat[\centering Comparison of different methods on the BSD100 dataset]{\label{eval_BSD100}\includegraphics[width=.9\linewidth]{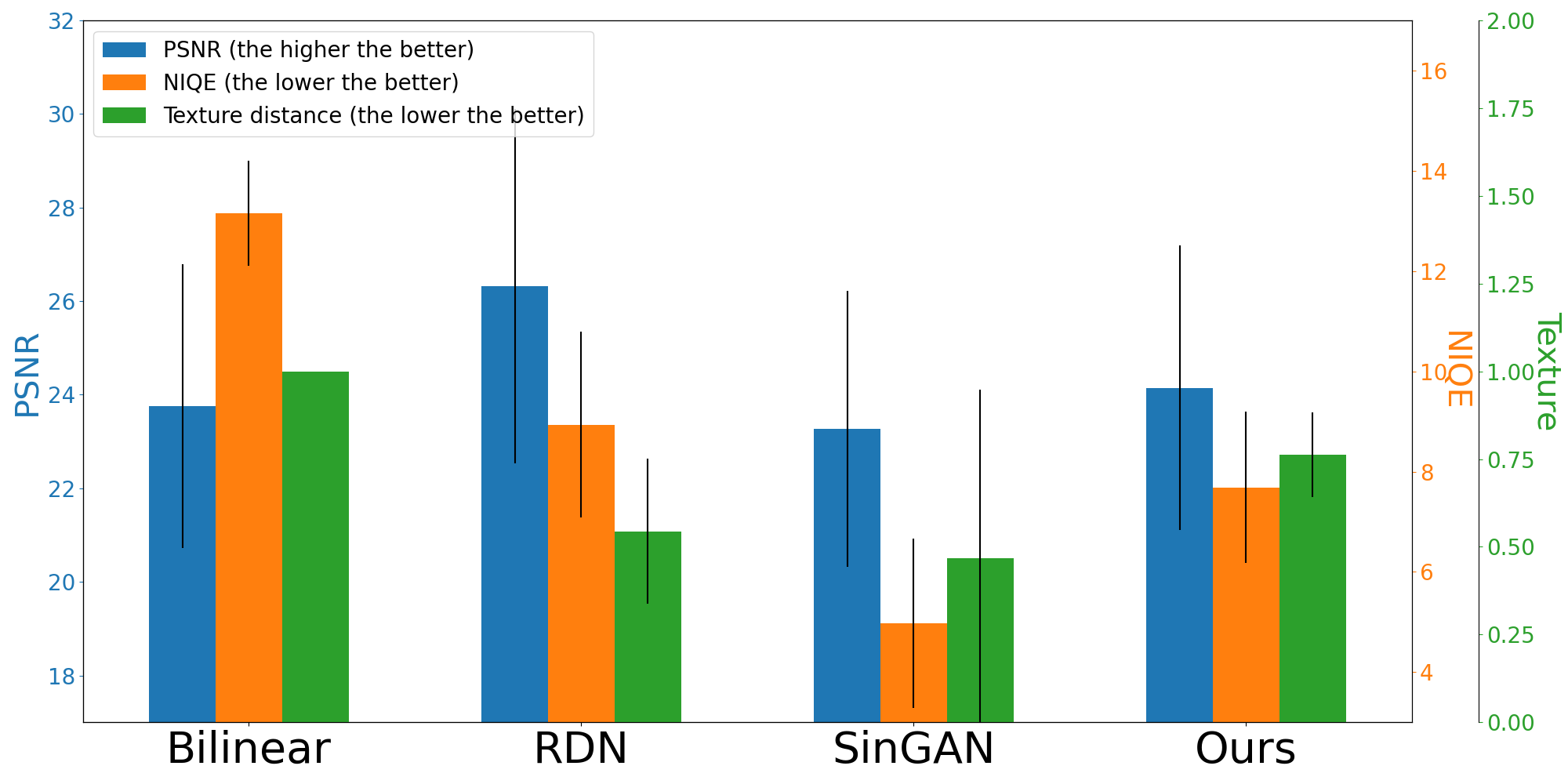}}
   
    \caption{\centering Evaluation on different datasets for our 3 metrics: PSNR, NIQE and texture distance} 
    \label{eval}
\end{figure}

\subsubsection{Synthetic dataset}
As a sanity check, we verify that our method performs well on colored noise images. We use images from our synthetic dataset, downsample them by a factor 4, then evaluate several methods when zooming of a factor 4. We emphasize that our method assumes a spectral slope of 1.7 for its spectral constraint, and these images are created to have the same slope of 1.7, thus giving a lot of prior information to our method which indeed outperforms the others in texture quality while maintaining a good PSNR (see Fig.\ref{sanity_check}). Apart from NIQE scores comparable to ours, SinGAN performs particularly poorly on this dataset.\\

\subsubsection{Cloud and BSD100 datasets}
On the BSD100 dataset (Fig. \ref{eval_BSD100}), SinGAN beats us in both perceptual quality metrics. This is logical since the natural images that form the BSD100 dataset are very different from colored noise like images. Our spectral prior is not valid on this dataset, which explains this degradation.\\
On the cloud dataset however (Fig. \ref{eval_cloud}), we report better texture and PSNR results than SinGAN, while beating RDN and bilinear interpolation on both perceptual metrics.\\

Let us note that our PSNR results (Fig. \ref{eval}) are always worse than RDN and slightly worse or comparable with the bilinear interpolation. We do not claim to compete in terms of PSNR with an external method such as RDN optimized for PSNR performance. We aim at respecting our constraints, and by doing that we yield good perceptual results, while maintaining acceptable PSNR scores.

\subsection{Respect of the constraints}
We verify on the Cloud dataset that the output of our method also respects the constraints we have set with three metrics, closely related to these constraints:
\begin{itemize}
\item \textbf{Color histogram distance} We compute a sliced transport distance between the color histograms of the ground truth and the output of the four evaluated methods.
\item \textbf{Reversibility error} We compute the $L_2$ distance between the low resolution image, and the blurred and downsampled SR output $w$: $\| D_f(w\ast g) - u \|^2$
\item \textbf{Straight slope error} We compute the standard error in the least square regression of the log-log graph (radial mean of Fourier modulus against frequency $f$). It measures how linearly the log-modulus decays in log-frequency. In other words, it measures how colored noise-like an image is. 
\end{itemize}
The results are reported in TABLE II.

\begin{table}[h!]
\centering
\scalebox{0.85}{
 \begin{tabular}{|c|c|c|c|c|} 
 \hline
  & Bilinear & RDN & SinGAN & Ours \\ 
 \hline
 Color histogram distance & $39\pm 28$ & $\mathbf{17} \pm \mathbf{3}$ & $938 \pm 2202$ & $47 \pm 40$ \\ 
 \hline
 Reversibility error ($\cdot 10^{-4}$) & $1.2\pm0.4$ & $\mathbf{0.5}\pm \mathbf{0.2}$ & $20.4\pm 49.8$& $0.6\pm0.3$ \\
 \hline
 Straight slope error ($\cdot 10^{-3}$) & $8.0\pm 1.2$ & $8.6\pm 1.3$ & $4.6 \pm 1.2$ & $\mathbf{3.3}\pm \mathbf{0.9}$\\

 \hline
 \end{tabular}}
 \label{constraints_check}
 \caption{Evaluation of the respect of the 3 constraints for all 4 methods on the Cloud dataset}
\end{table}
Keeping in mind that our Histogram constraint uses as reference the low resolution image (whose histogram differs slightly from the high resolution ground truth), we see that our method performs reasonably in this color evaluation.\\
Color artifacts and lack of reversibility from SinGAN are particularly salient in this table with very poor measures of color histogram distance and reversibility error.
Adding constraints allows to significantly reduce these problems.\\
Our methods, as expected, yields the best performance in terms of spectral slope linearity, followed by SinGAN. 

\section{Conclusion}
In this paper, we have introduced an internal SR method which preserves statistics that are related to the specificity of SR tasks and to decreasing property of the Fourier spectrum. The interest of our approach is demonstrated on comparison results with SinGAN, which we improve in terms of PSNR and texture fidelity on colored noise images datasets.
\section*{Acknowledgments}
This research was partially supported by the Defence Innovation Agency (Pierrick Chatillon PhD grant) and the ANR-19-CE40-0005 MISTIC project.

\bibliographystyle{IEEEtran}
\bibliography{IEEEabrv,root.bib}

\end{document}